\documentclass[12pt]{article}
\usepackage{amsfonts,epsfig}

\def\beq{\begin{eqnarray}}
\def\eeq{\end{eqnarray}}
\def\bea{\begin{eqnarray}}
\def\eea{\end{eqnarray}}
\begin{document}
\title{ASTROPHYSICAL OBSERVATIONS OF EARLY UNIVERSE PHASE TRANSITIONS} 
\author{Leonard S. Kisslinger$^\dagger$\\
    Department of Physics, Carnegie Mellon University, Pittsburgh, PA 15213}
\date{}
\maketitle

\section{ Review of CMBR Correlations}

Until about 380,000 years after the ``big bang'' the temperature of the 
universe was higher than atomic binding energies and atoms were ionized.
At that time the temperature fell to about 0.25 ev, electrons bound to
atomic nuclei, and the scattering of electromagnetic radiation from the early
periods of the universe almost ceased. We now observe this radiation after
this last scattering, called the Cosmic Microwave Background Radiation (CMBR).

\subsection{Evolution of the Early Universe}

During the period before the last scattering there were many important
astrophysical processes. For todays purposes we classify them by the time, t,
and temperature, T, at the time as:
\vspace{2mm}

{\bf Very early: inflation, strings, textures, etc.}

{\bf t$\simeq 10^{-11}s$; T $\simeq 100 GeV \simeq$ Higgs Mass: The
early universe Electroweak Phase Transition (EWPT)}

{\bf t$\simeq 10^{-5}s$; T $\simeq$ 300 MeV: the early universe QCD Chiral 
Phase Transition, or quark hadron phase transitiion(QHPT) 
(at T $\simeq$ 150 MeV).}

{\bf t$\simeq$ 380,000 years: Last Scattering, CMBR free, carries 
information about cosmological processes in the early universe.}

   Our main interest in the present work is to to
find if  the EW or QH phase transitions can produce observable effects
in the CMBR radiation or other astrophysical observations. 
 
\subsection{Formalism for CMBR Correlatons}

The starting point is to make a T=Temperature map using blackbody spectrum
with the microwave telescope in the $\hat{n}$ direction.  Expanding 
T($\hat{n}$) in spherical harmonics
\beq
\label{T(n)}
  \frac{T(\hat{n})}{T_0} = 1 + \sum_{lm} a^T_{lm}Y_{lm}(\hat{n}) \;.
\eeq
From $T_0 = 2.73 K^o \rightarrow$ we know that the big bang occured
 $\sim$ 14 billion years ago. The temperature fluctuations, $\Delta T$, are
given by $a^T_{lm}$. They are random: $<a^T_{lm}> = 0$.

\subsubsection{Temperature and Polarization Correlations}

The CMBR correlations are defined in terms of the Stokes parameters
for a plane electromagnetic wave (see, e.g., Jackson's text on Classical 
Electrodynamics): $I \sim$ intensity or T, Q and U $\sim$ E and B
type of polarization, and V $\sim$ circular polarization. The Stokes parameter
V does not have to be considered. The diagonal correlations, 
with $<a^T_{l'm'}a^T_{lm}>= C^{TT}_l \delta_{ll'}\delta_{mm'}$, are 
\beq
\label{CMBcor}
    <\frac{\Delta T(\hat{q}_1)}{T_0}\frac{\Delta T(\hat{q}_2)}{T_0}> &=& 
 \sum_{l}\frac{2 l +1}{4\pi} C^{TT}_l P_l(cos\alpha) \\
   <\frac{Q(\hat{q}_1)}{T_0}\frac{Q(\hat{q}_2)}{T_0}>
  &=& \sum_{l}\frac{2 l +1}{4\pi} C^{EE}_l P_l(cos\alpha) \\
  <\frac{U(\hat{q}_1)}{T_0}\frac{U(\hat{q}_2)}{T_0}> 
       &=& \sum_{l}\frac{2 l +1}{4\pi} C^{BB}_l P_l(cos\alpha) \; .
\eeq
{\bf The Temperature and polarizations correlations, $C^{TT}_l, C^{EE}_l, 
C^{BB}_l$, as well as the cross correlations, are the key observations of 
CMBR experiments.}

\subsubsection{Cosmology and CMBR Measurements}

Consider some cosmological event, such as inflation, EWPT, or QHPT. These
processes can give rise to seeds. One must evolve these seeds to the last
scattering time. We briefly review this, using the Hu-White\cite{huwhite}
formalism

   Let us defines the vector $\vec{T}(\Theta,Q+iU,Q-iU), \Theta=\Delta T/T$,
with the three components constructed from the Stokes parameters.
The Boltzman equation to evolve to last scattering time is

\hspace{2cm} $\frac{d\vec{T}}{d\eta} = C[T] + G[h_{\mu\nu}]$\\
with $\eta = t/a(t)$ = conformal time, a(t)= cosmic scale,
$g_{\mu\nu}=a(t)^2[g_{\mu\nu}^0 +h_{\mu\nu}$ = metric tensor,
$h_{\mu\nu}\rightarrow$  metric perturbations,
$C[T]\rightarrow$ Compton scattering, and
$G[h_{\mu\nu}]\rightarrow$ gravitational (metric) fluctuations.
 
   To find correlations $C^{TT}_l,C^{EE}_l,C^{BB}_l$
one expands in angular modes,
inserts $\Theta$, Q, and U in the Boltzman equation, and does the
 angular integrals. The Boltzman equation takes seed from time of 
cosmological events to last scattering time to obtain $C^{XY}_l$
The solutions to Boltzman Eq. at conformal time $\eta_0$ for the
angular projections of the Stokes functions are
\beq
\label{bolzT}
  \frac{\Theta^m_l(\eta_0,\vec{k})}{2l+1} &=& \int_{0}^{\eta_0} d\eta
\frac{g(\eta_0,\eta)}{\dot{\tau}(\eta)}\sum_{l'}S^{m}_{l'}j^{l',m}_l
[k(\eta_0-\eta)] \; ,
\eeq
\beq
\label{bolzB}
  \frac{B^m_l(\eta_0,\vec{k})}{2l+1} &=& -\sqrt{6} \int_{0}^{\eta_0} d\eta
g(\tau_0(\eta),\tau(\eta))P^m \beta^m_l [k(\eta_0-\eta)] \; ,
\eeq
and a similar expression for $E^m_l$, with $j^{l',m}_l,\beta^m_l$ related
to Bessel functions, with $\tau(\eta)$ the optical depth, $g(\tau_0,\tau)$
is the visibility function,and $S^m_l(\eta),P^m(\eta)$ are the sources of 
T,E,B fluctuations. 

$C^{X,X'}$ fluctuations are obtained from integrals over products of
$\Theta^m_l, E^m_l, B^m_l$ at $\eta_0$.

\subsection{CMBR Measurements, Very Early Univerwse Cosmological Predictions}

A recent history of CMBR measurements after {\bf COBE}, $l \leq 200$ is:

{\bf BOOMERANG} Baloon telescope, 1998, A.E. Lang et al, Phys. Rev. D63 
(2001)04200, Resolution 10 arcmin. $50\leq l \leq 800$

{\bf MAXIMA} Baloon telescope, 1998, S. Hanney et al, Ap.J. Lett. 545 ('00)5.
Resolution 10 arcmin  $35\leq l \leq 785$

{\bf ACBAR} Uses {\bf VIPER} (CMU), Antartica

{\bf DASI} Interferometer, Antartica

{\bf WMAP} Wilkinson Microwave Anisotropy Probe, C.L. Bennet et.al., 
Astrophys. J. Suppl. 148 (2003) 1.

The first important thing that we learn from CMBR measurements in recent
years is that the
{\bf First (acoustic) peak at $l \simeq 200 \rightarrow \Omega \simeq 1.0$,
  FLAT UNIVERSE}.
From the recent WMAP experiment {\bf TE polarizations have been seen}. There
are prospects that BB correlations, which are most important for our present
work (as we discuss in the next section), will be observed

As mentioned above, cosmological theorists must find effects that can
observed. Predictions of CMBR correlations from defect models and inflation
models have been made (see \cite{seljak,kamion}). For our present purposes
the most important results are for BB polarization, which we discuss in
the next section.

\section{QCD Phase Transition}

Starting from the QCD Lagrangian
\beq
\label{glue}
  \mathcal{L}^{glue} & = & -\frac{1}{4} G \cdot G
\eeq
\beq
\label{G}
    G_{\mu\nu}^n & = & \partial_\mu A_\nu -  \partial_\nu A_\mu
-i g [A_\mu,A_\nu] \nonumber \\
    A_\mu & = & A_\mu^n \lambda^n/2
\eeq
with $\lambda^n$ the eight SU(3) Gell-Mann matrices, use the 
instanton model, which gives for the QCD gluonic Lagrangian for
instantons of size $\rho$
\beq
\label{linst}
  \frac{1}{4} G^{inst} \cdot G^{inst}  & = & 48 \frac{\rho^4}{(x^2+\rho^2)^4}
\; .
\eeq
\subsection{Bubble Surface Tension in Instanton-Like Model}

  In order to investigate bubble collisions that could lead to observations
one must develope a good model for the bubble wall. For the QHPT we take
guidance from the instanton model\cite{b't} to calculate the surface 
tension\cite{lsk1}.
In the instanton model, a SU(2) Euclidean space model,the energy density
 $T^{00}$ = 0. In Minkowski space one finds
\beq
\label{Tinst}
   T^{00(inst)} &=& 3 (2)^5 \big( \frac{\rho}{x^2 + \rho^2} \big)^4 \; .
\eeq
The bubble wall surface tension for one instanton is
\beq
  \sigma^{inst}(T=0) &=& \int dx T^{00}(x,0,0) \nonumber \\
                     &=& \frac{30 \pi}{\rho^3} \; .
\eeq
Use lattice gauge picture-double peak and the $n(T_c) = 0.5 n(T=0)$
= 0.0004 from the liquid instanton model\cite{ss98} we find
\beq
          \sigma(T_c)^{instanton \; model} &=& .013 T_c^3 \\ \nonumber
          \sigma(lattice) &\simeq& .015 T_c^3 \; ,
\eeq
quite good agreement with the lattice\cite{bkp} calculation.

\subsection{Collisions of Instanton-like Bubbles}

 \begin{figure}[ht]
\centerline{\psfig{figure=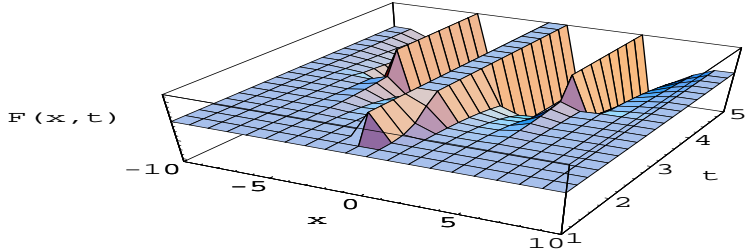,height=4cm,width=10cm}}
\caption{F(x,t) for the two bubbles with instanton-like form}
\end{figure}

   From the classical theory of bubble collisions, when two bubbles
collide an interior wall with the same surface tension as the original
bubble walls is formed. For the QHPT the our picture would have a 
short-lived QCD gluonic instanton-like wall. To test this we\cite{lsk2}
look for solutions to the equations of motion with the purely gluonic
Lagrangian, Eq.(~\ref{glue}) in 1+1 Minkowski space, with the gauge condition
$\partial_\mu A^a_\mu=0$. I will give a very brief description as the 
details of this are discussed by Mikkel Johnson in his talk at this Workshop. 

Using the instanton-like 
ansatz, with $\eta_{a\mu\nu}$ defined in Ref.~ \cite{b't},
\beq\label{A_inst}
   A^{a}_\mu(x) &=& \frac{2}{g}\eta_{a\mu\nu}x^\nu F(x^2) \equiv
 \eta_{a\mu\nu} W^\nu \; . 
\eeq
The equation of motion for F is
\beq
\label{eominst}
   \partial^2 F & = & -\frac{2}{x}\partial_x F -12 F^2 +8 (x^2-t^2)F^3 \;.
\eeq
Typical results showing the interior gluonic wall forming are shown in
Fig.1. This is the physical basis for our magnetic wall
and CMBR polarization correlations discussed in the next section.

\section{QHPT, Magnetic Walls, and CMBR Correlations}

In this section we discuss the magnetic wall which would form from
QCD bubble collisions during the QHPT, and how this could provide the
seed for observable BB polarization correlations in the CMBR\cite{lsk3}
Starting from the electromagnetic interaction
\beq
\label{Lint}
     {\cal L}^{int} & = & -e \bar{\Psi} \gamma^\mu A^{em}_\mu \Psi,
\eeq
where $\Psi$ is the nucleon field operator, in Ref~\cite{lsk3} a magnetic wall
was derived with the form
\beq
\label{wall}
  {\bf B}_W({\bf x}) &=& B_W e^{-b^2(x^2 + y^2)} e^{-M_n^2 z^2} \; ,
\eeq
 $b^{-1} =d_H \simeq$ few km, $M_n^{-1} \simeq 0.2 fm$, and  
$B_W \simeq \frac{3 e}{14 \pi} \Lambda_{QCD} \simeq  10^{17}$ Gauss within 
the wall. The compton scattering from this wall provides the seed,
$P^2$ in Eq.(\ref{bolzB}), from which the solution of the Boltzman equation
and the formalism described in Sec.I gives for the BB polarization 
correlation: 
\beq
\label{cbbf}
    C^{BB}_l & \simeq & 4.25\times10^{-8} l^2
\eeq
This is illustrated in Fig.2.
\begin{figure}[ht]
\begin{center}
\psfig{figure=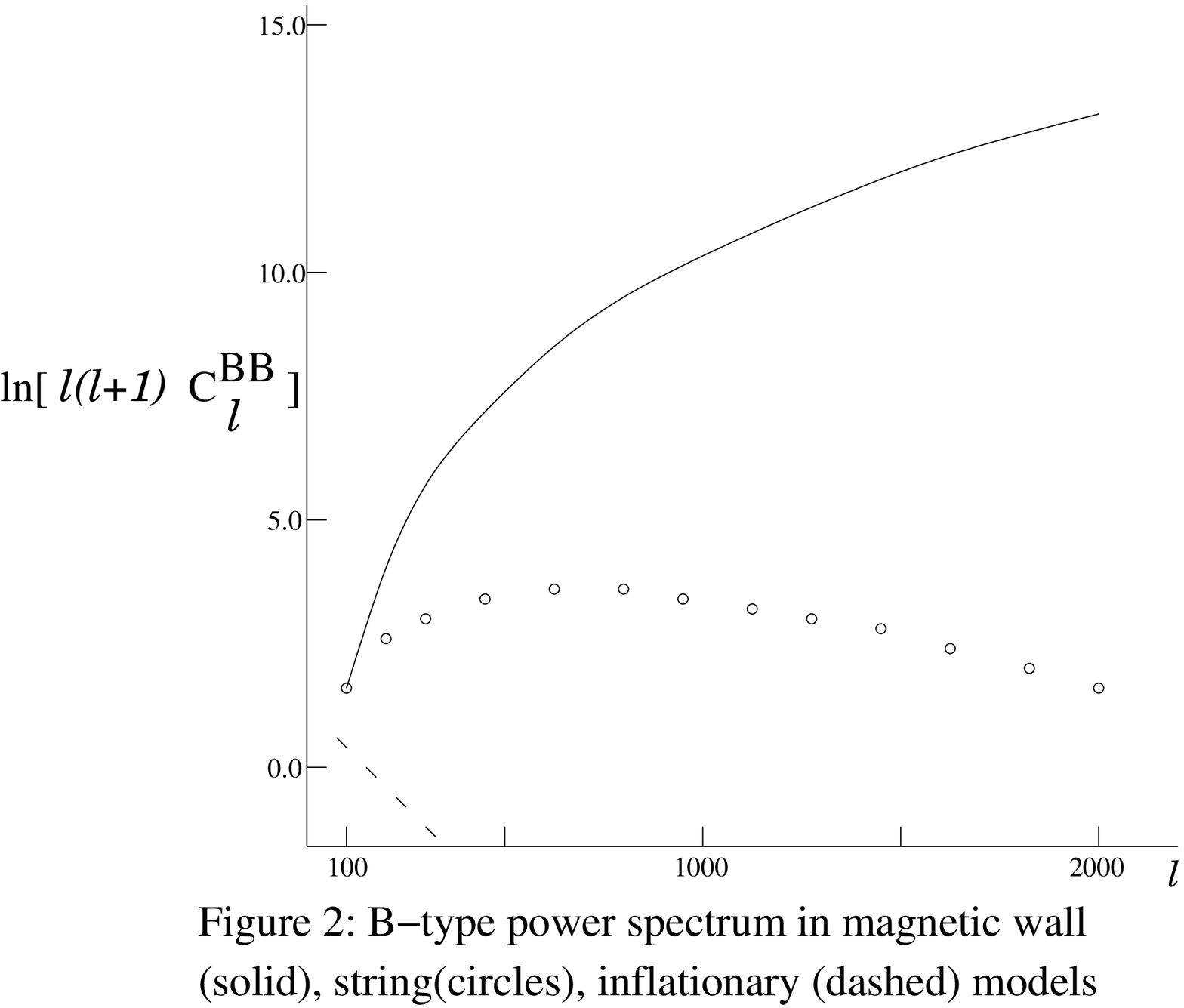,height=7cm,width=10cm}
\end{center}
\end{figure}
 Note that for large $l$ we find BB CMBR correlations which can be tested
as they are different from very early universe models.
 
  In Ref~\cite{lsk3} a calculation of metric fluctuations the $C^{BB}_l$
arising from metric fluctuation, gravity wave effects were also estimated.

\section{Electroweak Phase Transition}
 
   At a time t$\simeq 10^{11}$s T$\simeq$ 100 GeV. At that time the
EWPT occured, and the Higgs and gauge bosons acquired their masses.
In recent years it has been recognized\cite{kaj} that with the standard
Weinberg-Salam Model at finite T there is no first order phase transition,
and no EW bubble formation. Although even a crossover EWPT is important
for baryogenesis, there would be no observable cosmological effects.

   With a minimal supersymmetric model a first order EWPT is possible.
See, e.g., Ref~\cite{l96}. After integrating out the second Higgs and
all supersymmetric partners except the stop, the Lagrangian is of the form
\beq
\label{L}
  {\cal L}^{EW} & = & {\cal L}^{WS} + {\cal L}^{stop} \nonumber \\
  {\cal L}^{stop} & = & |D_\mu \Phi_t|^2 +m_t^2 |\Phi_t|^2 \nonumber \\
   D_\mu &=& i\partial_{\mu} -\frac{g}{2} \tau \cdot W_\mu
 - \frac{g'}{2}B_\mu \, ,
\eeq 
where the $W^i,W^3,B$, with i = (1,2),are the EW gauge fields, and 
$\Phi_t$ is the Stop field, with $\Phi_L,\Phi_R$. The e.o.m. are 
obtained from $\delta \int d^4 {\cal L}^{EW} = 0$
The solution of the e.o.m.is a PISETAL project (Choi, Henley, Hwang, Johnson, 
Walawalkar,and Kisslinger). Our main objective is to derive the magnetic 
fields generated in EWPT bubble collisions to find seeds for galactic and 
extra-galactic magnetic fields, an unsolved problem.

\section{Conclusions}
 
\hspace{5mm}{\bf The QCD chiral phase transition might lead to large-scale 
gluonic walls.}
\vspace{2mm}
 
{\bf If gluonic walls are metastable, magnetic walls would form.} 
\vspace{2mm}

{\bf Large-scale magnetic walls would lead to observable CMBR polarization
correlations and density fluctuations for $ l > 1000$.}
\vspace{2mm}

{\bf For the electroweak phase transition one needs the MSSM or other 
extensions of the WSM. Observable Magnetic fields in galaxy structure might 
result.}
\vspace{3mm}

   $^\dagger$ This work was supported in part by NSF grant PHY-00070888.


\begin{thebibliography}{99}
\bibitem{huwhite}W. Hu and M. White, Phys. Rev. {\bf D 56}, 596 (1997). 
\bibitem{seljak}U. Seljak, U-L. Pen and N. Turnok, Phys. Rev. Lett.{\bf 79}, 
1615 (1997).
\bibitem{kamion}M. Kamionkowski and A. Kosowsky, Phys. Rev. {\bf D 57}, 
685 (1998).
\bibitem{b't}A.A. Belavin et. al., Phys. Lett. {\bf 59}, 85 (1975);
G. 't Hooft, Phys. Rev. {\bf 14}, 3432 (1976).
\bibitem{lsk1}L.S. Kisslinger, hep-ph/0202159 (2002).
\bibitem{ss98}T. Schaffer and E.V. Shuryak,  Phys. Rev. Rev Mod. Phys.
{\bf 70}, 323 (1998).
\bibitem{bkp}B. Beinlich, F. Karsch and A. Peikert, Phys. Lett. {\bf B 390},
 268 (1997).
\bibitem{lsk2}M.B. Johnson, H-M. Choi and L.S. Kisslinger, Nucl. Phys.
 {bf A 729}, 729 (2003).
\bibitem{lsk3}L.S. Kisslinger, Phys. Rev. {\bf D 68}, 043516 (2003).
\bibitem{kaj}K. Kajantie, et. al. Phys. Rev. Lett.{\bf 14}, 2887 (1996).
\bibitem{l96}M. Laine, Nuc. Phys {\bf B481}, 43 (1996).

\end{thebibliography}
\end{document}